# In-field Remote Fingerprint Authentication using Human Body Communication and On-Hub Analytics

Debayan Das, Shovan Maity, Baibhab Chatterjee & Shreyas Sen, *Senior Member, IEEE*
School of Electrical and Computer Engineering, Purdue University, USA. {das60, shreyas}@purdue.edu

*Abstract*— In this emerging data-driven world, secure and ubiquitous authentication mechanisms are necessary prior to any confidential information delivery. Biometric authentication has been widely adopted as it provides a unique and non-transferable solution for user authentication. In this article, the authors envision the need for an in-field, remote and on-demand authentication system for a highly mobile and tactical environment, such as critical information delivery to soldiers in a battlefield. Fingerprint-based in-field biometric authentication combined with the conventional password-based techniques would ensure strong security of critical information delivery. The proposed in-field fingerprint authentication system involves: (i) wearable fingerprint sensor, (ii) template extraction (TE) algorithm, (iii) data encryption, (iv) on-body and long-range communications, all of which are subject to energy constraints due to the requirement of small form-factor wearable devices. This paper explores the design space and provides an optimized solution for resource allocation to enable energy-efficient in-field fingerprint-based authentication. Using Human Body Communication (HBC) for the on-body data transfer along with the analytics (TE algorithm) on the hub allows for the maximum lifetime of the energy-sparse sensor. A custom-built hardware prototype using COTS components demonstrates the feasibility of the in-field fingerprint authentication framework.

*Index Terms*— Authentication, biometrics, Human Body Communication, wearable sensors, fingerprint identification.

## I. INTRODUCTION

Rapid advancement of technology in the last few decades has led to the escalating storage and access of private information in electronic devices. Hence, the need for a secure and robust authentication mechanism is extremely critical, prior to accessing any confidential information [1].

### A. Vision: In-field Wearable Biometric Authentication

In this work, the authors envision the need for an in-field authentication system in dynamic, hostile and resource-constrained environments, like the military battlefields. Authentication is necessary to detect any insider threats, and to ensure that no information is being leaked to outsiders.

Biometric authentication provides a more reliable, non-transferrable and convenient solution to user authentication [2]. In addition, a soldier in the battlefield would prefer to spend minimal time to authenticate himself. Moreover, logging into a mobile device or a radio, physically by hand can delay combat actions. Hence, a wearable biometric-based technique is necessary to provide a seamless in-field authentication.

### B. System Overview & Related Work

The authentication technique for a military battlefield needs to be in-field, that is, the person authenticating the military personnel on the battlefield, is a few miles away at the base station (cloud). Several remote biometric authentication schemes have been proposed to address security weaknesses

This work was supported in part by the NSF CRII Award No. CNS 17-19235, NSF SaTC Award No. CNS 16-57455 and the Air Force Office of Scientific Research YIP Award (FA9550-17-1-0450). The experimental protocols involving human subjects have been approved by the Purdue Institutional Review Board (IRB Protocol #1610018370).

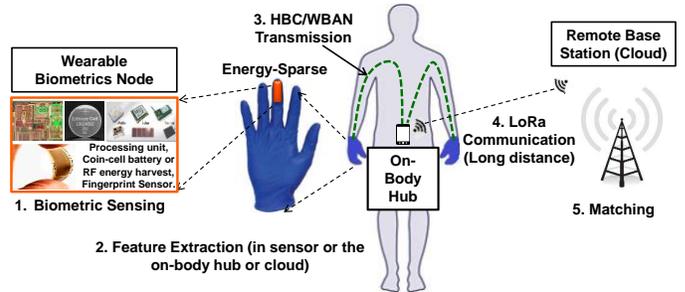

Figure 1: Vision: In-field Wearable Biometric Authentication System: Biometric data collected by the wearable fingerprint sensor (embedded in the glove) is communicated to the hub using on-body communication. The hub sends the encrypted biometric data to the remote base station (cloud) for matching with the stored database. In this work, in-sensor computation (TE) and communication (Wireless, HBC) modalities are analyzed, highlighting the trade-offs and best choice along with hardware demonstration.

[3], [4]. However, these traditional remote authentication schemes are not in-field and hence do not have any energy limitation, contrary to in-field authentication with energy-sparse wearable physiological sensor nodes.

Figure 1 shows an overview of our proposed in-field authentication system. It comprises of (i) a wearable fingerprint sensor embedded in the glove, for biometric sensing, (ii) template or feature-extraction (TE) algorithm, (iii) short-range on-body communication using wireless body area network (WBAN) or HBC [5], [6], (iv) an on-body hub or aggregator, (v) long-range communication using LoRa protocol [7], (vi) the remote base station to match the received fingerprint data with the stored database. As seen from Figure 1, the fingerprint image captured by the wearable sensor node is sent to the on-body hub using HBC or WBAN, which transmits the data to a distant base station, using LoRa.

### C. Contribution

Specific contributions of this paper are:

- Envisioned the need for a remote, in-field, wearable biometric authentication system for dynamic and tactical environments, such as a military battlefield.
- Efficient on-body communication and Edge-Analytics are explored to provide an optimized solution for maximizing lifetime of both the sensor node and the on-body hub simultaneously. On-body HBC transmission is chosen to reduce the energy per bit and Edge-Analytics (high-accuracy TE) is allocated to the on-body hub in order to maximize the system lifetime.
- Compared to [8], this work for the first time, utilizes *on-hub analytics* along with HBC to develop an efficient body area network. The *application demonstration* includes a proof-of-concept in-field fingerprint authentication system prototype in hardware, using capacitive HBC for the on-body link, template extraction on the hub, followed by LoRa communication for the hub-tower link.

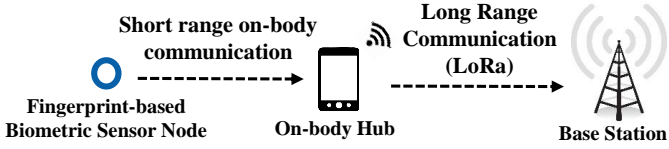

| | Sensor Node | On-body communication | Hub/Aggregator | Cloud |
|---|---|---|---|---|
| (a) | TE | WBAN | No TE + LoRa | Matching |
| (b) | | HBC | | |
| (c) | No TE | WBAN | TE + LoRa | Matching |
| (d) | | HBC | | |
| (e) | No TE | WBAN | No TE + LoRa | TE + Matching |
| (f) | | HBC | | |

Figure 2: Six prominent Possiblities of the proposed Remote Biometric System. TE denotes *template extraction*. (a, b): TE in the sensor and on-body transmission using HBC & WBAN respectively, (c, d): TE in the aggregator and on-body transmission using HBC & WBAN respectively, (e, f): TE in the remote BS, and on-body transmission using HBC & WBAN respectively. In all cases, LoRa communication is used from the aggregator to the cloud.

## II. SYSTEM COMPONENTS OF THE PROPOSED IN-FIELD BIOMETRIC AUTHENTICATION SYSTEM

The in-field fingerprint authentication system comprises of biometric sensing, data compression, encryption and communication. As shown in Figure 2, the six prominent possibilities for resource allocation in the proposed in-field authentication system involve (i) selection of a wearable fingerprint sensor, (ii) choice and allocation of the TE algorithm, (iii) selection of the on-body communication technique. This section develops the optimization framework to perform resource partitioning for the authentication system to maximize lifetime of the energy-sparse sensor and the hub.

### A. Biometric Sensing

Fingerprint image acquisition technologies typically involve optical and solid-state capacitive sensors [9], [10]. As a wearable device, capacitive sensors are more effective as it can be built on-chip and consumes lower power ($22.3\ nJ/capture$), compared to an optical sensor ($66\ mJ/capture$), although it provides less accuracy in terms of the image capture.

### B. Template Extraction Algorithms: A Comparative analysis

The accuracy of fingerprint matching highly depends on the choice of the template extraction (TE) algorithm. In order to select the desired algorithm, two sets – (i) a high-accuracy, and (ii) a lightweight minutiae-based TE algorithm, are implemented and compared in terms of their accuracy, energy consumption, and the compression ratio (CR) [10]. Figure 3 shows the comparison of the two TE algorithms against high quality (Figure 3(a)) and blurred fingerprint (Figure 3(f)) images taken from the FVC2002 database. As seen from Figure 3(b, c, g, h), the sophisticated high-accuracy algorithm determines all the ridge endings (marked in red) and the bifurcations (marked in blue) accurately. On the other hand, the lightweight TE algorithm (Figure 3(d, e, i, j)) performs poorly in terms of accuracy. Hence, the high-accuracy TE algorithm is desirable for reliable fingerprint matching even at the expense of higher energy ($2.94\ J/capture$) with $CR = 227\times$, resulting in compressed minutiae data of $\frac{40032}{227} \approx 176\ Bytes$ (Table 1).

### C. Communication: Intra-body and Long-range

The fingerprint sensor node captures the image, and needs to transmit it to the cloud for matching. The energy-sparse sensor node transmits the data to the on-body hub using intra-body communication (WBAN/HBC). WBAN consumes an energy of $10\ nJ/bit$ [11], whereas HBC consumes $\sim 79\ pJ/bit$ [5]. The hub then wirelessly transmits the data to the cloud server using long-range (LoRa) communication protocol ($\sim 68\ \mu J/bit$) [7]. This high energy requirement for LoRa transmission further justifies the need for a hub, since the edge sensor is energy-constrained and hence cannot directly transmit data to the cloud.

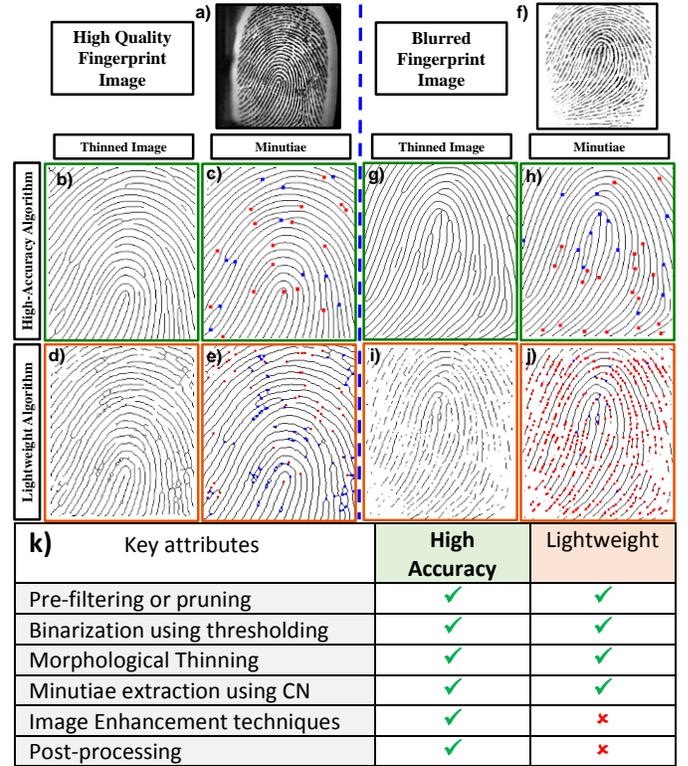

Figure 3: Accuracy comparison of the fingerprint extraction algorithms against high quality image (a-e) and blurred image (f-j) respectively. (a) High quality fingerprint image, (b, c) Thinned image & the extracted minutiae using the high-accuracy (sophisticated) algorithm for the good image, (d, e) Thinned image & the extracted minutiae using the lightweight algorithm for the good image, (f) Blurred fingerprint image, (g, h) Thinned image & the extracted minutiae using the high-accuracy algorithm for the blurred image, (i, j) Thinned image & the extracted minutiae using the lightweight algorithm for the blurred image, (k) table showing the comparison of the key attributes between the two algorithms.

Unlike WBAN, HBC signals are mostly contained within the human body making it more secure, and hence encryption energy for HBC is not considered in our analysis.

## III. PROPOSED BIOMETRIC SYSTEM: OPTIMAL ALLOCATION OF RESOURCES & COMMUNICATION NETWORKS

In this section, the six different possibilities (Figure 2) for the in-field fingerprint-based biometric authentication system, using an optical and a capacitive sensor are quantitatively analyzed. Table 1 shows the necessary parameters used for analyzing the framework [5], [7], [9], [11] - [13]. The energy availabilities in each of the sensor nodes is shown in Table 1.

The total energy consumption for the wearable sensor node as well as the hub, for one time user authentication, is given as,

$$E_{total_{edge/hub}} = E_{capture} + E_{comp} + E_{comm} + E_{encr} \quad (1)$$

where, $E_{capture}$ denotes the image capture energy, $E_{comp}$ denotes the energy consumed by the TE algorithm for one request, $E_{comm}$ is the communication energy to transmit the fingerprint data. In case of the hub, $E_{capture} = 0$. The lifetime (number of retries) of a node in the system is given as,

| Parameters | Values |
|---|---|
| Energy per bit: WBAN | 10 $nJ/bit$ [11] |
| Energy per bit: HBC | 79 $pJ/bit$ [5] |
| Energy per bit: LoRa (500 m) | 68 $\mu J/bit$ [7] |
| Energy per bit: Encryption ($E_{encr}$) | 100 $pJ/bit$ [12] |
| Fingerprint image capture size | 40032 Bytes or 320.256 Kbits |
| Extracted Template (minutiae data) size | 176 Bytes or 1408 bits [#] |
| Image Capture Energy consumption ($E_{capture}$) | 22.3 $nJ$ (capacitive sensor) [9]  66 $mJ$ (optical sensor) |
| TE algorithm Energy consumption ($E_{comp}$) | 2.94 $J$ (high-accuracy) |
| Energy available: Sensor node | RF Energy Harvest: 1 $\mu W - hr$ [13]  Coin cell: 100 $mW - hr$ |
| Energy available: On-body Hub | Li Ion Poly Battery: 4.5 $W - hr$ |
| Energy available: Base Station | Cloud (no limits): 100 $KW - hr$ |

[#] Minutiae data size for the good quality image using high-accuracy algorithm.
Table 1: Parameters used in the proposed biometric framework.

$$R_{edge/hub} = \frac{E_{av_{edge/hub}}}{E_{total_{edge/hub}}} \qquad (2)$$

where, $E_{av}$ denotes the total energy availability for that node as shown in Table 1. Note that, for a chargeable coin-cell battery, $R$ refers to the number of retries per charge; whereas for the RF energy harvesting, $R$ denotes the number of retries in an hour.

### A. Data Compression (TE) in the fingerprint sensor

With the compression (TE) in the sensor (capacitive or optical), WBAN or HBC can be used to transmit the extracted minutiae data from the edge to the hub. Using Eqs. 1, 2 and from Figure 4(a, e), we see that the compression energy dominates and the lifetime of the coin-cell battery operated sensor is 119 and 122 retries (per charge of the battery), for the optical and capacitive sensors respectively (Figure 4(b, f)). However, it can also be seen that *RF energy harvesting does not support TE in the sensor*.

### B. No data compression (TE) in the fingerprint sensor

In this scenario, the TE can be performed either in the on-body hub, or the remote cloud. Now, using Eq. 1 and as shown in Figure 4(c, d), the image capture energy dominates for the optical sensor. Although it supports ~5K retries per charge (Figure 4(d)) of the coin-cell battery for both HBC and WBAN, the optical sensor cannot support energy harvested edge sensor.

Using a *capacitive sensor*, the capture energy reduces drastically (Figure 4(g, h)), and it is able to support an energy-harvested sensor with both WBAN and HBC. However, *HBC provides > 128× higher lifetime* (~142 retries per hour) compared to WBAN (~1 retry per hour).

Now, we analyze the lifetime of the hub and optimally allocate the compression (TE) algorithm in the hub or in cloud.

*1) Data Compression (TE) in the Hub:* In this scenario, the hub compresses the data and sends it to the cloud using LoRa. However, the hub of the system might serve as the common node for multiple edge devices, and hence we assume the available energy for hub dedicated to the fingerprint authentication is $E'_{av_{hub}} \sim E_{av_{hub}} * 10\%$. Using Eqs. 1, 2 and using LoRa ($E_{comm_{lora}} \alpha D^2$) for a distance ($D$) of 1 $Km$, $R_{hub} = 483$ per charge of the lithium ion battery.

*2) Data Compression (TE) in the Cloud:* Here, the hub needs to transmit the raw fingerprint data to the distant base station (cloud) using LoRa. As the LoRa communication energy is high, for $D = 1 Km$, the lifetime of the hub reduces to $R_{hub} = 18$ per charge (using Eqs. 1, 2), which is significantly low.

Hence, the *high-accuracy TE (analytics) in the hub is an optimal allocation*.

## IV. EXPERIMENTAL RESULTS: ON-HUB ANALYTICS AND HBC-BASED EFFICIENT FINGERPRINT AUTHENTICATION

In this section, as a proof of concept, we demonstrate a custom hardware prototype of the fingerprint authentication system using capacitive HBC for the on-body communication, and LoRa radios for hub-cloud communication.

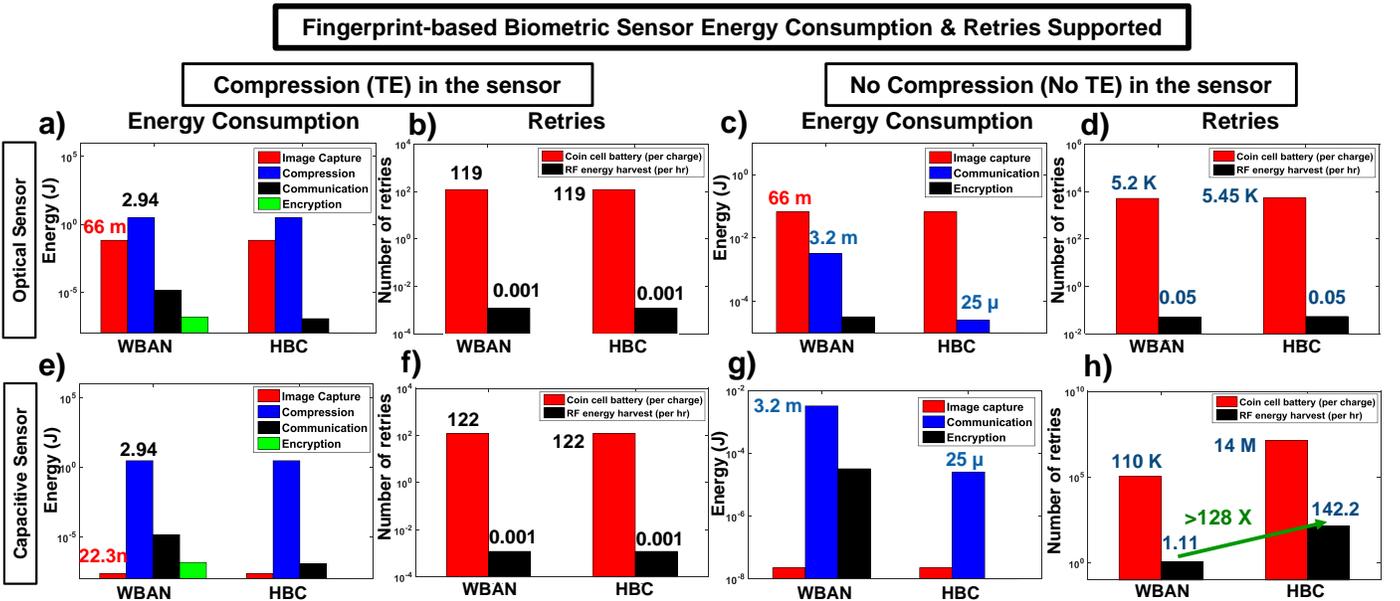

Figure 4: Sensor Node Energy Consumption and number of requests (retries) supported. (a, b) Optical sensor, with the template extraction (TE) algorithm in the sensor, (c, d) Optical sensor, without TE in the sensor, (e, f) Capacitive sensor, with the template extraction (TE) algorithm in the sensor, (g, h) Capacitive sensor, without TE in the sensor.

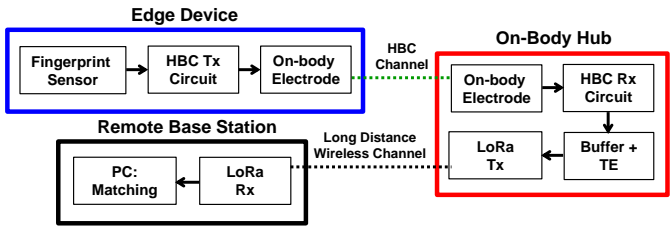

Figure 5: Block Diagram of the *Hardware prototype* for in-field fingerprint authentication. Tx and Rx represent transmitter and receiever respectively. For *system-level feasibility demonstration*, a battery-powered optical sensor is used at the edge. SD Card is used to buffer the incoming data stream at the hub. Finally, PC was used to match the fingerprint with the stored database.

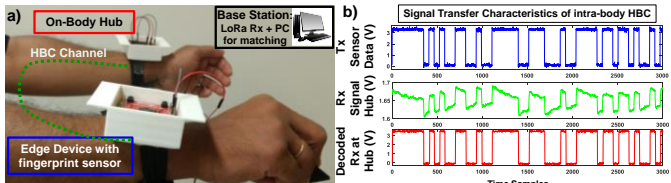

Figure 6: (a) Data transfer between edge and hub using HBC. (b) Fingerprint data from edge (blue) is transmitted using HBC to the hub. Green plot is the received signal (distorted) by the hub, which then decodes the data (red).

### A. Components and Set-up

The authentication system nodes – biometric sensor, on-body hub, and the receiving base station, are built using commercial off-the-shelf (COTS) components, as shown in Figure 5, Figure 6(a). For the sake of demonstration, all the nodes are battery powered to emulate the scenario of a wearable device. The biometric sensor node consists of an optical fingerprint sensor (ZFM-20), an ATmega1280-based microcontroller for fingerprint image processing, and custom designed copper electrode-bands for coupling the signals from the microcontroller in to the human body (HBC). An optical fingerprint sensor is used due to its wide availability, and for the purpose of demonstration of the in-field authentication system. The on-body hub consists of a similar interface copper electrode band to receive the data transmitted through the body (HBC), and another microcontroller (HBC receiver) to receive, compress (analytics) and transmit the data using LoRa radio (SX1231: 915 MHz ISM band). For reliable and continuous transmission of the fingerprint data, it is buffered in the hub using a SD card, before transmitting through the LoRa radio. Finally, the remote base station (Figure 5, Figure 6(a)) consists of a LoRa receiver, and a computer to match the received fingerprint data with the stored database.

### B. Intra-body HBC: Key Techniques

In HBC, the transmitter and the receiver do not share a common ground, and completes the return path using capacitance to ground [13]. Hence, only AC signals can be transmitted (Figure 6(b), blue waveform) through the human body. The human body acts as a single wire, and hence the fingerprint data must be sent serially. However, standard serial transmission protocols like the UART that needs a common ground, are unreliable, and hence, a custom protocol is used for broadband data transmission. The received signals are attenuated due to the weak return path [13], [14]. (Figure 6(b), green waveform). To eliminate the 60 Hz power supply noise picked up by the human body, acting as an antenna, a high-pass bias circuit is used. In addition, to reduce the effect of in-band interferences, an integrator is used to increase the opening of the eye diagram [15], so that the received signal can be reconstructed correctly (Figure 6(b), red waveform).

**RF energy-harvested sensor node**

| Retries per hour | TE in the sensor | | TE in the hub | |
|---|---|---|---|---|
| | WBAN | HBC | WBAN | **HBC** |
| Optical sensor | 0.001 | 0.001 | 0.05 | 0.05 |
| | ✗ | ✗ | ✗ | ✗ |
| **Capacitive sensor** | 0.001 | 0.001 | 1.11 | 142.2 |
| | ✗ | ✗ | ✗ | ✓✓ |

Table 2: Summary of outcomes – Fingerprint Sensor Lifetime (retries per hour) with *RF energy-harvesting*: Only HBC for the on-body communication can support >100 retries per hour, with capacitive sensing and the analytics (high-accuracy TE algorithm) in the on-body hub.

## V. CONCLUSION

This work envisions the need for in-field biometric authentication in a resource-constrained environment, like the military battlefield. Efficient on-body communication and Edge-Analytics were explored to provide an optimized solution for maximizing lifetime of both the sensor node and the on-body hub simultaneously. As summarized in Table 2 (derived from Figure 4), the most optimal allocation involves a capacitive sensor, HBC for the on-body communication, placement of high-accuracy compression algorithm (analytics) in the on-body hub, along with LoRa communication for the long distance hub-cloud link, and finally fingerprint template matching at the cloud server. This work, for the first time, uses *on-hub analytics* with HBC, and shows a proof-of-concept demonstration of the in-field fingerprint authentication system with a custom-built hardware prototype.